\documentstyle[psfig]{mn}
\def\snr{G309.2--00.6}
\def\src{ATCA~J134649--625235}
\newcommand\HI{H\,{\sc i}}
\newcommand\HII{H\,{\sc ii}}
\newcommand\kms{km~s$^{-1}$}

\begin{document}
\label{firstpage}
\title[\snr\ and jets in supernova remnants]
{\snr\ and jets in supernova remnants}
\author[B. Gaensler, A. Green \& R. Manchester]
{B. M. Gaensler$^{1,2}$\thanks{E-mail: b.gaensler@physics.usyd.edu.au
(BMG); a.green@physics.usyd.edu.au (AJG); rmanches@atnf.csiro.au 
(RNM)}\thanks{Address after Sep 1998: Center for Space Research, Massachusetts
Institute of Technology, Cambridge, MA 02139, USA}, 
A. J. Green$^1$\raisebox{.8ex}{$\star$} and R. N. Manchester$^2$\raisebox{.8ex}{$\star$} \\
$^1$Astrophysics Department, School of
Physics A29, University of Sydney, NSW 2006, Australia \\
$^2$Australia Telescope National Facility, CSIRO, PO Box 76,
Epping, NSW 2121, Australia}

\pagerange{\pageref{firstpage}--\pageref{lastpage}}
\pubyear{1998}

\maketitle
\begin{abstract}

We present Australia Telescope Compact Array observations of the
supernova remnant (SNR) \snr. In a 1.3-GHz continuum image the remnant
appears as a near-circular shell, but with two
brightened and distorted arcs of emission on opposite sides.  
\HI\ absorption against the
SNR yields a distance in the range 5.4 to 14.1~kpc, corresponding to
an age $(1-20) \times 10^3$~yr.  

On the basis of the SNR's
morphology we argue that it is a younger
analogue of the W~50 / SS~433 system, and that its unusual appearance
is a result of opposed jets or outflows from a central source. A
jet-like feature and breaks in the shell can both be seen along the
axis of proposed outflow, providing further support for this
interpretation; the central source itself is not detected.  The SNR may
be interacting with the adjacent \HII\ region RCW~80 through an
extension of the proposed outflow beyond its shell.  This would put the
SNR at the lower limit of its distance range and would imply an age
$\la$4000~yr.  We consider other SNRs similar to \snr, and propose
remnants whose shells are affected by jets as one
of several classes of SNR from which the presence of a central source can
be inferred.

\end{abstract}


\section{Introduction}
\label{sec_g309_intro}

Radio observations of supernova remnants (SNRs) demonstrate a vast
range of shapes (e.g.\ Whiteoak \& Green 1996\nocite{wg96}).
While most SNRs have a distorted and complicated appearance reflecting
their interaction with an inhomogeneous interstellar medium (ISM), some
SNRs have striking symmetry properties which require other explanations
(e.g.\ Manchester 1987; Roger et al. 1988; R\'{o}\.{z}yczka et al.
1993; Gaensler 1998\nocite{man87,rmk+88,rtfb93,gae98}).

\snr\ was first identified as a SNR on the basis of its non-thermal
spectrum \cite{gre74,ccg75}. Subsequent higher resolution observations
\cite{cmw81,kc87,wg96} have shown a distorted shell with two opposed,
symmetric bright ends, and a weak compact source in the interior.
Continuing a programme to study unusual southern SNRs (Gaensler,
Manchester \& Green 1998a\nocite{gmg98}, hereafter Paper~I), we 
present high resolution 1.3-GHz continuum and \HI\ absorption
observations of \snr, as well as observations of the region in
H$\alpha$, in X-rays and in the near-infrared. In
Section~\ref{sec_g309_obs} we briefly describe our observations and
analysis, before presenting our results in
Section~\ref{sec_g309_results}. In Section~\ref{sec_g309_discuss} the morphology
of SNR~\snr\ is discussed, and is compared to that of other SNRs.

\section{Radio observations and reduction}
\label{sec_g309_obs}

Radio observations were carried out with the Australia Telescope
Compact Array (ATCA; Frater, Brooks \& Whiteoak 1992\nocite{fbw92}), a
six-element synthesis telescope near Narrabri, New South Wales.  Three
different array configurations were used, as shown in
Table~\ref{tab_g309_observations}.  Observations were made simultaneously in the
radio continuum (centre frequency 1.344~GHz) and in the \HI\ line
(centre frequency 1.420~GHz, channel separation 0.83~\kms) towards a
pointing centre RA (J2000) $13^{\rm h}46^{\rm m}35^{\rm s}$, Dec
(J2000) $-62\degr53\arcmin48\arcsec$.  All other details of the
observations, calibration and analysis are as in Paper~I.

\begin{table}
\caption{ATCA observations of \snr.}
\label{tab_g309_observations}
\begin{tabular}{cccc} 
Date    & Array  & Maximum & Time on \\
	& Configuration & Baseline (m) & Source (h) \\ \hline
1996 Jan 18 & 0.75C & 750 & 13 \\
1996 Feb 03 & 0.75B & 765 & 12 \\
1996 Feb 25 &  1.5C & 1485 & 13 \\ \hline
\end{tabular}
\end{table}

\begin{table}
\caption{Observational and derived parameters for \snr.}
\label{tab_g309_snr}
\begin{tabular}{lc} \hline 
Resolution & $24\farcs1 \times 22\farcs7$, PA 14$^{\circ}$ \\
rms noise in image  & 160 (Stokes $I$) \\
\hspace{1cm} ($\mu$Jy~beam$^{-1}$)$^a$ & 50 (Stokes $V$) \\
Geometric centre ($\alpha$, $\delta$; J2000) &
$13^{\rm h}46^{\rm m}37^{\rm s}$ --62\degr53\arcmin\ \\
Geometric centre ($l$, $b$) & 309\fdg17 --00\fdg68 ($l$, $b$) \\
Diameter (arcmin) & $12 \times 10$, PA --40\degr (shell) \\
                  & $14 \times 6$, PA +45\degr (ears) \\
Flux density at 0.4~GHz (Jy)$^{a,d}$ & $10\pm1$ \\      
Flux density at 0.8~GHz (Jy)$^{b,d}$ & $6.0\pm0.6$ \\
Flux density at 1.3~GHz (Jy)$^{c,d}$ & $5.2\pm0.2$\\
Spectral index & $-0.53\pm0.09$ \\ \hline
\end{tabular}
\\
\footnotesize
$^a$Mills Cross data \cite{gre74} \\
$^b$MOST data \cite{wg96} \\
$^c$ATCA data (this paper) \\
$^d$~1~jansky (Jy) $= 10^{-26}$ W m$^{-2}$ ; errors of 10 per cent have
been assumed in Mills Cross and MOST data
\end{table}

\section{Results}
\label{sec_g309_results}

\subsection{Total Intensity}
\label{sec_g309_results_i}

Fig~1 shows total intensity images of \snr, while
Fig~\ref{fig_g309_primary_beam} shows an image of the entire
field. Properties of sources of note are given in 
Table~\ref{tab_g309_sourcelist}. Source 3 corresponds to the
\HII\ region G309.548--0.737 \cite{hcs79,ch87b}.

\begin{figure*}
\label{fig_g309_snr}
\vspace{10cm}
\caption{ATCA images of SNR~\snr\ at 1.3~GHz: (a) grey
scale representation, ranging from --1 to 15~mJy~beam$^{-1}$; (b)
contours with levels at 2, 5, 10, 15
and 20~mJy~beam$^{-1}$; (c) grey scale with
range --0.1 to 1.5~mJy~beam$^{-1}$, shown to emphasise faint structure.
The grey scales have not been corrected for the ATCA primary beam
response, in order to give uniform noise across the image. The FWHM of
the Gaussian restoring beam is shown at lower right of each panel.}
\end{figure*}

\begin{figure*}
\begin{minipage}{140mm}
\centerline{\psfig{file=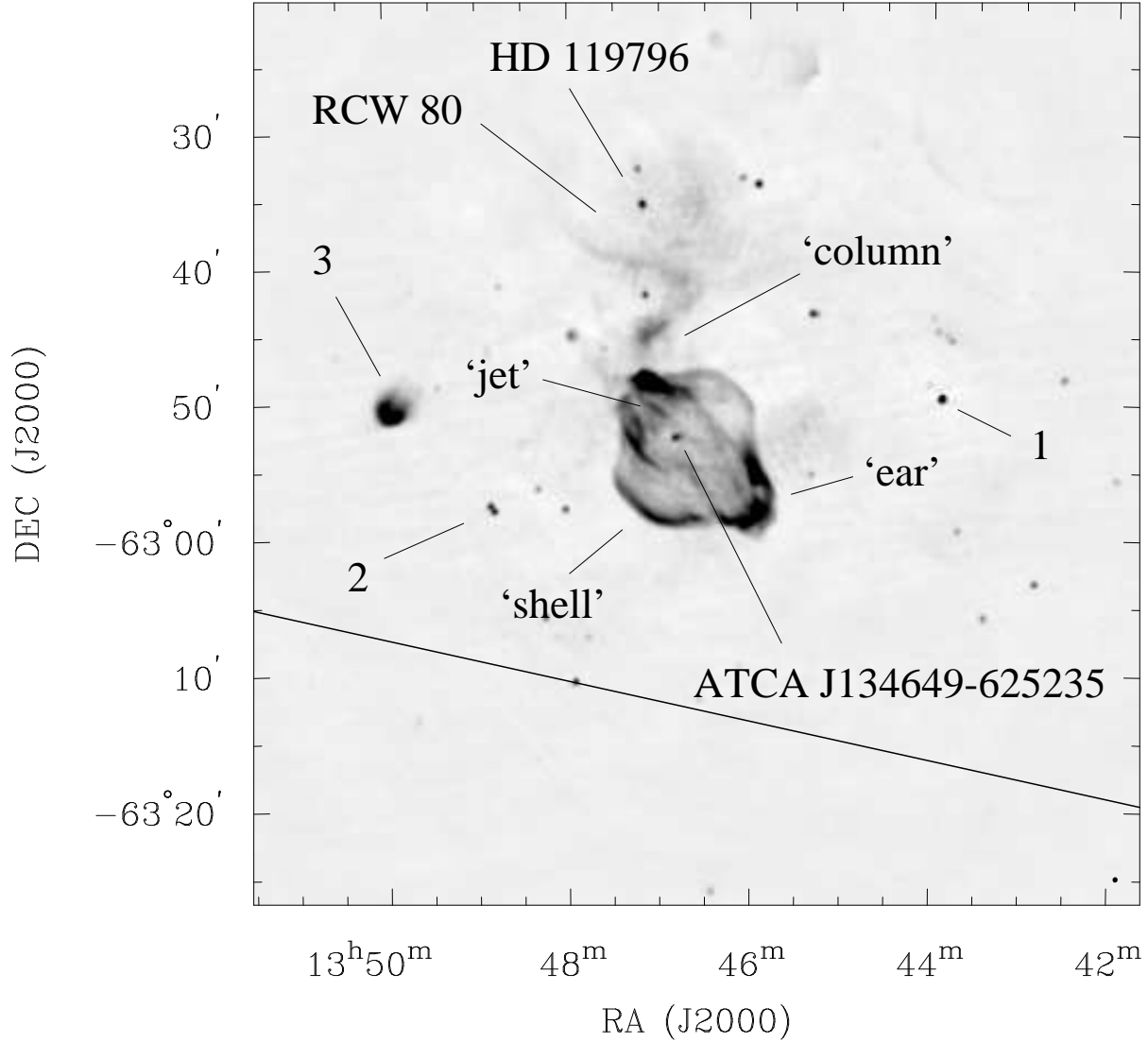,width=16cm}}
\caption{A total intensity image of the field surrounding SNR~\snr. The
image has not been corrected for the ATCA primary beam response.
Sources listed in Table~\ref{tab_g309_sourcelist} are indicated, as well as
some of the main features associated with the SNR.  The diagonal line
running across the image corresponds to a Galactic latitude
$b=-1\degr$.}
\label{fig_g309_primary_beam}
\end{minipage}
\end{figure*}

\begin{table*}
\begin{minipage}{160mm}
\caption{Selected sources in the vicinity of SNR~\snr.}
\label{tab_g309_sourcelist}
\begin{tabular}{cccccl} 
Source  & \multicolumn{2}{c}{Position} &
$S_{\rm 1.3\,GHz}$ & Spectral index$^a$ & Other names \\
& Equatorial (J2000) & Galactic 
& (Jy) & ($\alpha$, $S_{\nu} \propto \nu^{\alpha}$) &  \\ \hline
1      & 13$^{\rm h}$43$^{\rm m}$53$^{\rm s}$ --62\degr49\arcmin41\arcsec & 
    308\fdg88 --00\fdg56 & 0.08 & $-0.7$   \\
central source     & 13$^{\rm h}$46$^{\rm m}$49$^{\rm s}$ 
--62\degr52\arcmin35\arcsec & 309\fdg20 --00\fdg68 & 0.02  & +0.6$\pm$0.6 &
\src \\
HD~119796 & 13$^{\rm h}$47$^{\rm m}$10$^{\rm s}$
--62\degr35\arcmin20\arcsec
      & 309\fdg30 --00\fdg41 &  0.06 & 0.0 & HR 5171, IRAS 13436--6220 \\ 
RCW~80    & 13$^{\rm h}$47$^{\rm m}$ --62\degr38\arcmin
      & 309\fdg3 --00\fdg4 &  0.7 & +0.3$\pm$0.5 & Gum 48d \\ 
column  & 13$^{\rm h}$47$^{\rm m}$ --62\degr45\arcmin
      & 309\fdg2 --00\fdg6 &  0.6 & --0.6$\pm$0.5 & \\
jet    & 13$^{\rm h}$47$^{\rm m}$ 
--62\degr50\arcmin & 309\fdg2 --00\fdg6 & 0.02  & --- \\
2     & 13$^{\rm h}$48$^{\rm m}$50$^{\rm s}$ --62\degr57\arcmin51\arcsec & 
    309\fdg40 --00\fdg82 & 0.06 & $-1.0$ \\
3     & 13$^{\rm h}$50$^{\rm m}$ --62\degr50\arcmin
      & 309\fdg6 --00\fdg7 & 1.7 & $-0.1$ & G309.548--0.737, PMN J1349--6250 \\ 
\hline

\end{tabular}

$^a$ Calculated between 1.344~GHz (this paper) and 843~MHz \cite{gcl98}.  

\end{minipage}
\end{table*}
The remnant is comprised of two morphological components: firstly two
roughly circular arcs of emission to the south-east and north-west
which we subsequently refer to as the `shell', and secondly two bright,
sharply curved arcs to the south-west and north-east (the `ears'). The
two parts of each component are diametrically opposed with respect to
the SNR's geometric centre, and the two components are oriented at
position angles perpendicular to each other. Although no connecting
structure is apparent, the two arcs of the shell can be construed to form a
single circular ring. 

\begin{figure}
\centerline{\psfig{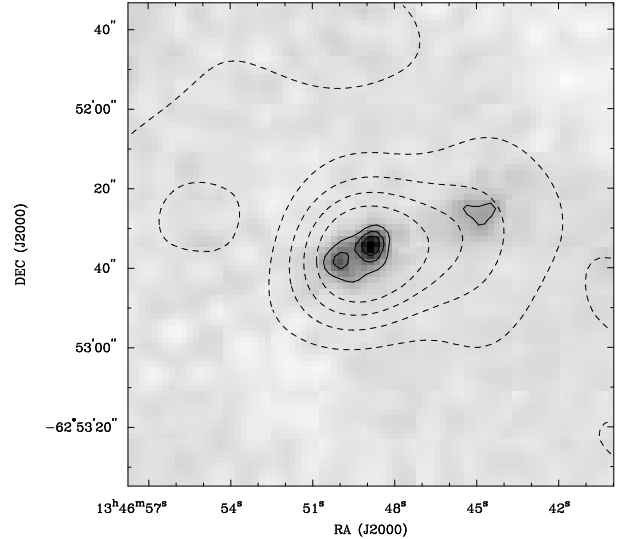}}
\caption{An image of the source \src\  within \snr, using all six
ATCA antennae. The greyscale runs from --0.5 to 4~mJy~beam$^{-1}$,
with solid contours at 1, 2, 3 and 4~mJy~beam$^{-1}$. The dashed
contours correspond to the lower resolution image shown in 
Fig~1(b), and are at levels of 2, 4, 6 and 8~mJy~beam$^{-1}$.}
\label{fig_g309_snr_central}
\end{figure}

Within the remnant, to the north-east of centre, is a slightly extended
source which we designate \src. At the available resolution ($\sim$20
arcsec -- see Table~\ref{tab_g309_snr}), this source
has a cometary
appearance, with a tail trailing out to the west.  In
Fig~\ref{fig_g309_snr_central} is shown an image of \src\
made using the observations in Table~\ref{tab_g309_observations}, but
including the sixth ATCA antenna, 3~km west of the track upon
which the other five antennae are stationed.  This increases the
maximum baseline to $\sim$5000~m, corresponding to a resolution of 5
arcsec. (Note that this gives a significant gap in the $u-v$ coverage
between 1500~m and 5000~m, and so is not appropriate for imaging the
entire remnant.) At this higher resolution, the source breaks into a
double source to the east, and a fainter extended source to the west.  

To the north of the remnant is a narrow column of emission. Extending
northwards from the SNR's north-eastern ear, this column bends around to the
west and then to the north again, before opening up into a broader
diffuse region forming a semi-circular arc.  The arc is coincident with
the H$\alpha$ nebula RCW~80 \cite{rcw60,gbg+88}, while within
it is a point source coincident with the position of 
the double star HD~119796 \cite{hsn71,hs85}.

Between \src\ and the north-eastern
ear, is a collimated feature extending away from the SNR's
centre, which we dub the `jet'.  There is a distinct break in the
emission from the north-eastern ear where it intersects the jet. The jet
and the break both lie along the symmetry axis defined by the two
ears.
The bright linear
component of the jet is 1\farcm5 long. At its south-west end the jet
abruptly fades, although a faint continuation can be seen extending to
the SNR's centre.  Beyond the break in the north-eastern ear is
faint emission extending 3 arcmin beyond the SNR.

The south-west ear also shows a break, although it is less distinct,
and does not lie along the symmetry axis. Faint emission which might
correspond to a less distinct and less collimated counterpart to
the jet is seen just within this ear.

Derived parameters for the SNR are given in Table~\ref{tab_g309_snr}, and
were determined by methods described in Paper~I\nocite{gmg98}.  Total
flux density measurements of \snr\ are shown in
Table~\ref{tab_g309_snr} -- we exclude single dish observations
\cite{dtg69,ccg75,dshj95}, which are confused by emission from RCW~80
to the north. We consequently compute a spectral index for SNR~\snr\ of
$\alpha = -0.53\pm 0.09$, where $S_\nu \propto \nu^\alpha$.  The
spectral index calculated here is somewhat steeper than previous
results ($\alpha = -0.37$; Clark et al. 1975a\nocite{ccg75}), probably
due to confusion with RCW~80.  We note that the largest spatial scale
sampled in our image is 17 arcmin, only slightly larger than the
remnant. Although our data include additional spacings corresponding to
scales of 13 and 10 arcmin, one could argue that we are missing some of
the SNR's flux density, and that its spectrum is consequently flatter
than that we have just determined. However, the ATCA flux density is
greater than that extrapolated from lower frequency data, and we argue that 
little flux is missing.  

Although \src, RCW~80 and the column feature are of low
surface brightness, we can put rough constraints on their spectra,
as listed in Table~\ref{tab_g309_sourcelist}. The lack of resolution in
the MOST image of Whiteoak \& Green \shortcite{wg96} 
prevents a calculation of the jet's 843~MHz flux
density and hence of its spectrum.

\subsection{Polarization}
\label{sec_g309_results_pol}

\subsubsection{SNR~\snr}
\label{sec_g309_results_pol_snr}

\begin{figure}
\centerline{\psfig{file=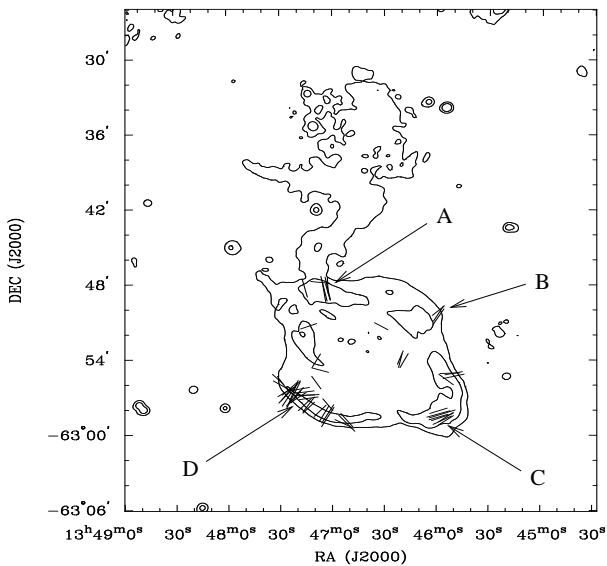,width=8cm,angle=270}}
\caption{Linearly polarized emission from \snr. The orientation of
vectors indicate the position angle of the electric field in an 8~MHz
channel centred on 1376~MHz. Lengths of vectors are proportional to the
surface brightness in linear polarization at that position, the longest
vector corresponding to $L=2.0$~mJy~beam$^{-1}$. Contours
representing total intensity are drawn at 2 and 10 mJy~beam$^{-1}$.
Labelled regions correspond to the plots in
Fig~\ref{fig_g309_snr_rm}.}
\label{fig_g309_snr_pol}
\end{figure}

Images of polarized emission were produced as in Paper~I, the effects
of bandwidth depolarization being minimised by imaging Stokes $Q$ and $U$
in 13 distinct frequency channels, forming $L=(Q^2+U^2)^{1/2}$ for
each channel, and then combining.
No circular
polarization was detected from \snr; linear polarization from
the SNR is shown in Fig~\ref{fig_g309_snr_pol}.  We put a lower limit of
75~mJy on its linear polarization, corresponding to an
overall fractional polarization of 1.4~per cent (instrumental
polarization at the field centre is negligible). However as a function of
position, the fractional polarization in some places reaches the
theoretical maximum of 70\%. Much of the SNR's outer edge is
polarized, correlating roughly with total intensity, with a rough
consistency in position angle over a given region. There is no
suggestion that the ear components have different polarimetric
properties from the shell. No polarization is detectable from 
\src\ or from the jet.

\begin{figure}
\centerline{\psfig{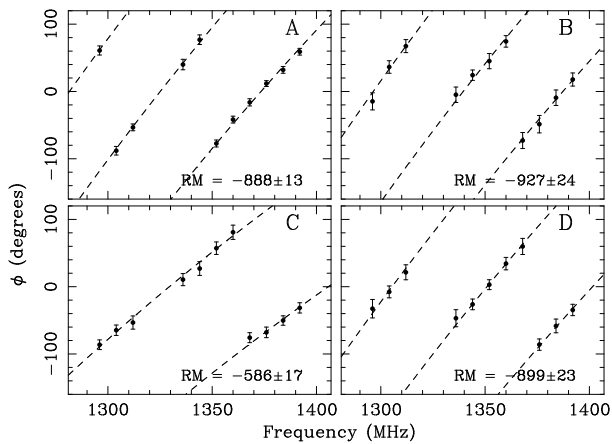}}
\caption{Faraday rotation across the observing band for four
regions of SNR~\snr\ indicated in Fig~\ref{fig_g309_snr_pol}.
The data correspond to position angles of the electric field
at 8~MHz intervals; gaps at 1320 and 1328~MHz are due to 
data corrupted by interference. The broken line represents
the best-fit curve of the form $\phi = \phi_0 + {\rm RM}\, c^2/\nu^2$;
the fitted rotation measures are given in rad~m$^{-2}$.}
\label{fig_g309_snr_rm}
\end{figure}

As demonstrated in Paper~I\nocite{gmg98}, the multiple channels
recorded in the ATCA's continuum mode can be used to derive a rotation
measure (RM) towards linearly polarized sources. In
Fig~\ref{fig_g309_snr_rm} we show the frequency dependence of
polarization position angle for four regions of the SNR. Towards the
two shell components and towards the north-eastern ear, we find a consistent
RM of  --930~rad~m$^{-2}$ with fluctuations (1~$\sigma$) of
60~rad~m$^{-2}$. The south-western ear has a distinctly different RM of
--570~rad~m$^{-2}$, with 20~rad~m$^{-2}$ fluctuations. As discussed in
Paper~I, the frequency range across which measurements have been made
is relatively small, resulting in large errors when
extrapolating the position angles to infinite frequency. Thus from
the current data we cannot determine the intrinsic orientation of
polarization vectors in the source.

\subsubsection{Other sources}
\label{sec_g309_results_pol_other}

No linear polarization is detected from RCW~80 or from the column,
either formally at the 5-$\sigma$ level, or by eye at any lower level.
Sources 1 and 3 are both 4 per cent linearly polarized
after correction for instrumental polarization. Source 2 and
HD~119796 are less than
3 per cent linearly polarized.  No circular polarization is detected from any
source in the field.

\subsection{\HI\ line}
\label{sec_g309_results_line}

As in Paper~I, we convert systemic LSR velocities into distances using
the best fitting model for Galactic rotation of Fich, Blitz \& Stark
\shortcite{fbs89}, and assume $\pm7$~\kms\ uncertainties in velocities.
We adopt a solar orbital velocity $\Theta_0 = 220$~\kms\ and a
distance to the Galactic Centre $R_0 = 8.5$~kpc, as recommended by Kerr
\& Lynden-Bell~\shortcite{klb86}. 

Our ATCA observations lack the $u-v$ data at short spacings required to
produce useful \HI\ emission spectra. Thus in considering
\HI\ absorption towards sources of interest, we compare with emission
seen towards the nearby ($\sim 2\degr$) SNR~G311.5--00.3
\cite{cmr+75}.  Their profile shows continuous emission from --60~\kms\
up to 0~\kms, then a strong peak at +35~\kms\ and weaker emission near
+100~\kms. Emission in the surveys of Jackson \shortcite{jac76} and
Kerr et al.  \shortcite{kbjk86} shows similar structure.

\subsubsection{Nearby sources}
\label{sec_g309_results_line_nearby}

\begin{figure*}
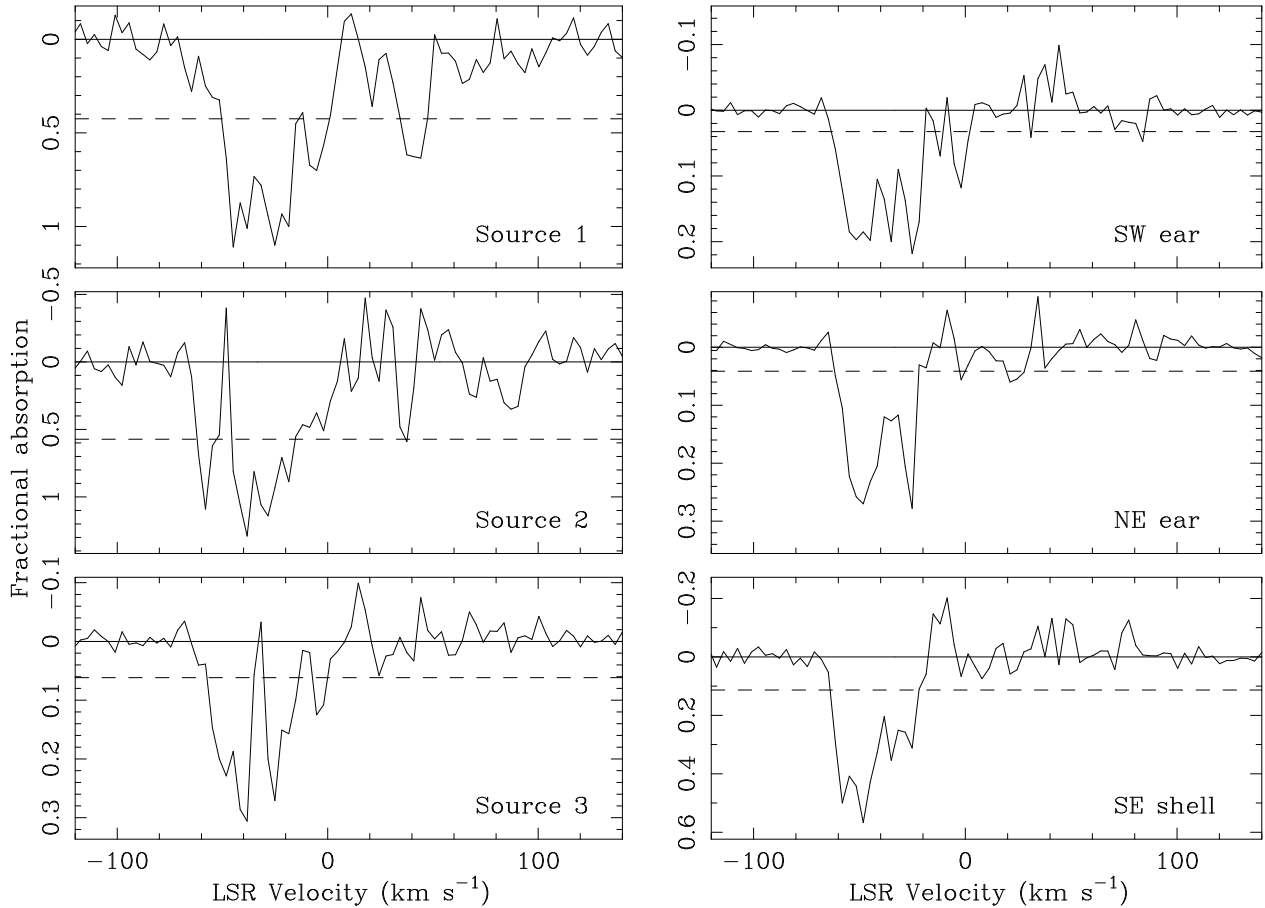

\begin{minipage}{160mm}
\centerline{\psfig{file=figure6a.eps,height=12cm}
\hspace{5mm}
\psfig{file=figure6b.eps,height=12cm}}
\caption{\HI\ absorption spectra towards sources 1, 2 and 3, and towards
three parts of the SNR. The dashed
line represents absorption at the 6-$\sigma$ level, where $\sigma$
is calculated from the signal in line-free channels.}
\label{fig_g309_hi}
\end{minipage}
\end{figure*}

Absorption was measured against sources 1, 2 and 3; the results
are shown in Fig~\ref{fig_g309_hi}.  Absorption towards source 1 is
seen down to --50~\kms, corresponding well to the observed tangent
velocity in this direction \cite{jac76,kbjk86}, and to that expected
from the rotation curve of Fich et al. \shortcite{fbs89}.  Absorption towards
source 1 is also seen at --5~\kms, and then again at +40~\kms.  These
features are similar to those seen in emission towards G311.5--00.3 as
described above. Towards source 2, significant absorption is seen at
negative velocities out to --60~\kms.  A weak feature is seen at
+40~\kms.  Absorption against source 3 is detected between --50 and
--40~\kms, then at --25~\kms\ and --5~\kms. No absorption is detected at
positive velocities. Other sources in the image, including the column,
HD~119796 and RCW~80, were too weak or too far from the phase
centre to obtain useful absorption against.

\subsubsection{SNR~\snr}
\label{sec_g309_results_line_snr}

Useful absorption spectra were obtained against three parts of the
remnant: towards each ear, and towards the south-east shell
component, as shown in Fig~\ref{fig_g309_hi}.

Absorption towards the south-western ear is seen between --50 and
--25~\kms, possibly at --10 and then at 0~\kms. No
convincing absorption is seen at positive velocities.  Towards the
north-eastern ear, absorption is detected at --50 and \mbox{--25}~\kms.  Along
the south-eastern shell component, we see absorption only between --60 and
--25~\kms.

Examination of individual velocity channels clearly shows the outline
of the SNR in absorption, demonstrating that it is indeed
continuum emission from the SNR which is being absorbed and not
background \HI\ features (cf.\ Landecker, Roger \& Dewdney
1982\nocite{lrd82}). In any case, the brightness temperature of the SNR
in the relevant regions is 20--25~K, significantly brighter than
typical \HI\ emission features.

\subsection{Other wavelengths}
\label{sec_g309_results_othwave}

\begin{figure*}
\vspace{10cm}
\caption{Multi-wavelength observations of the region surrounding \snr.
The grey scales correspond to:
(a) H$\alpha$ emission;
(b) X-ray ({\em ROSAT}\ PSPC) emission in the energy range 0.5--2.0~keV,
smoothed with a $25''$ Gaussian;
(c) 60~$\mu$m ({\em IRAS} HIRES) emission.
Contours correspond to 1.3~GHz ATCA data at the levels
of 2 and 10~mJy~beam$^{-1}$.}
\label{fig_g309_ir_x_ha}
\end{figure*}

\subsubsection{Optical observations}

A field containing \snr\ was observed by A.J. Walker on 1997 Apr 30, as
part of a wide-field H$\alpha$ survey of the Southern Galactic Plane
being carried out at Siding Spring Observatory \cite{bbw98}.  Three
10~min exposures were made with a 400~mm, f/4.5 Nikkor-Q lens, through
a filter centred at 657.0~nm with a width of 1.5~nm. The detector was a
2048$\times$2048 CCD with a resolution of 12\arcsec~pixel$^{-1}$.  The
images were bias and dark subtracted, flat-fielded and then combined
with a median filter.  A final image, without continuum subtraction, is
shown in Fig~7(a).

RCW~80 is clearly delineated in H$\alpha$, having a similar arc-like
appearance to its radio morphology. As in the radio, HD~119796 is visible at
the centre of this region. No H$\alpha$ emission from the SNR
or from the column to its north is apparent.  
The open cluster NGC~5281 is visible
in the remnant's interior, at RA (J2000) $13^{\rm h}46^{\rm
m}30^{\rm s}$, Dec (J2000) $-62\degr55\arcmin$. Its distance
has been estimated to be 1.3~kpc \cite{mv73}.

\subsubsection{X-ray observations}
\label{sec_g309_other_xrays}

\snr\ was observed by the {\em ROSAT}\ Position Sensitive Proportional
Counter (PSPC) in 1992 Aug for 3931~sec; the data are now available
from the {\em ROSAT}\ public archive. In the energy range 0.1--0.4~keV,
only a slight enhancement above the background is seen towards the
SNR.  However in Fig ~7(b) is shown emission in the range 0.4--2.4~keV,
where the compact X-ray source 1WGA~J1346.5--6255 (White, Giommi \&
Angelini 1994a,b\nocite{wga94a,wga94b}) can be seen near the remnant's
centre, and weak diffuse emission can be associated with the rest of
the SNR. No significant emission can be seen along the column or from
RCW~80.

\subsubsection{Infrared observations}
\label{sec_g309_results_othwave_ir}

An {\em IRAS} 60~$\mu$m image of the region, processed with the HIRES
algorithm \cite{afm90}, is shown in Fig~7(c). Strong IR
emission is coincident with RCW~80 and with HD~119796 ($=$~IRAS~13436--6220), 
but no obvious emission can be
associated with the SNR or with the column to its north. The point
source on the north-western edge of the SNR is IRAS~13428--6232.

\section{Discussion}
\label{sec_g309_discuss}

\subsection{Field sources}

\HI\ absorption at +40~\kms\ towards source 1 puts it at a distance
of at least 14~kpc. Although no absorption is seen at +100~\kms\
corresponding to emission seen towards G311.5--00.3, this may
be due to the small angular size of absorbing clouds at this
large distance. Source 1's spectral index and low level of linear
polarization is consistent with it being a background radio source. 
Similar arguments apply to source 2, which in addition has the
characteristic double-lobed morphology of a radio galaxy.

Source 3 (G309.548--00.737) shows absorption out to the tangent point.
Using a recombination line velocity towards it of --43~\kms\
\cite{ch87b}, we can thus (within the uncertainties) put this source at
the tangent point, corresponding to a distance 5.4$\pm$1.6~kpc. The
source's morphology and flat spectrum are in accord with its
identification as an \HII\ region. The low level of linear polarization
we detect from it is not, however. Either there is a previously
unidentified synchrotron component to its emission, or the instrumental
polarimetric response of the ATCA is not being characterised correctly
at this large distance from the phase centre. Examination of the
position angle of polarized emission on a channel-by-channel basis
shows no evidence for Faraday rotation as is observed towards the SNR.
We thus consider instrumental effects to be a more
likely explanation. We note that instrumental response near the phase
centre is both minimal and well-characterised, and that polarimetry in
that part of the field is reliable (cf.\ Paper~I).

The infrared and H$\alpha$ emission associated with RCW~80 and its flat
spectrum are all consistent with it being a thermal \HII\ region. A
radial velocity $V_{\rm LSR} \approx
-48$~\kms\ \cite{gg70b,ap89b,ave97} puts it at the tangent point
($d=5.4\pm1.6$~kpc). One of the components of HD~119796 has been
proposed as the exciting star for RCW~80 \cite{gbg+88}, but its
photometric distance of 3.2~kpc \cite{hsn71} suggests that it may be
unrelated.

\subsection{SNR~\snr}
\label{sec_g309_discuss_snr}

\subsubsection{Physical Parameters}
\label{sec_g309_discuss_snr_params}

The three absorption spectra obtained towards the SNR are all
consistent, showing absorption out to the tangent point, but not at
positive velocities.  Thus we assign a lower velocity limit to the SNR
of $V_L = V_{\rm tangent} \approx -50$~\kms, and an upper limit $V_U =
+40$~\kms, corresponding to emission seen at this
velocity towards G311.5--00.3 \cite{cmr+75} and to absorption seen
against source 1 and weakly against source 2.

These lower and upper limits on the kinematic velocity correspond to
respective distances of 5.4$\pm$1.6~kpc and 14.1$\pm$0.7~kpc.  This
result immediately confirms NGC~5281
as an unrelated foreground object.  Using the shell
component to define the extent of the SNR, we thus find SNR~\snr\ to
have a diameter between 17 and 45~pc. Through arguments as in Paper~I,
this correponds to an age between 1000~yr (assuming the lower limit on
the distance and free expansion) and 20\,000~yr (assuming the upper
limit and an SNR in the adiabatic phase).  The low level of X-ray
emission towards the SNR suggests a line-of-sight column density of
$\sim10^{22}$~cm$^{-2}$ (cf.\  Hwang \& Markert 1994\nocite{hm94}, who
made marginal and non-detections of SNRs using comparable exposure
times), which for typical ISM densities is consistent with the distance
range determined using \HI\ absorption.  The Scutum-Crux arm of the
Galaxy lies in this direction at distances between 5 and 9~kpc
\cite{gg76,gbg+88}, which weakly favours a distance to \snr\  in the
lower half of the quoted range.

\subsubsection{Polarization}
\label{sec_g309_discuss_snr_pol}

\snr\ is only weakly polarized; as for
G296.8--00.3 in Paper~I, this is most simply explained in terms of
differential Faraday rotation within the beam, consistent with
the observed fluctuations in RM and typical of polarimetric
observations at this low frequency. 

The RM measured towards most of the SNR (regions A, B and D of
Fig~\ref{fig_g309_snr_pol}), when combined with a model for the Galactic
electron density distribution \cite{tc93}, corresponds to a mean
interstellar magnetic field along the line of sight in the range
1--9~$\mu$G, directed away from us.  Fig~\ref{fig_g309_snr_rm}
demonstrates a distinct difference in RM between the south-western ear
(region C) and the rest of the SNR, however. We now consider whether
this difference can be explained in terms of conditions internal to the
remnant. 

We first require a well-ordered field within the SNR, and indeed the
electric field vectors in this region are particularly uniform.  For a
distance to the SNR of $d = 10d_{10}$~kpc, the length of the line of
sight through the tip of the south-west ear is $l \la 9d_{10}$~pc.  We
assume a line-of-sight magnetic field within this region of $B =
10B_{10}$~$\mu$G, consistent with adiabatic compression of the
interstellar magnetic field by the SNR shock.  The resultant thermal
electron density required over the entire region is then at least
$5(B_{10}d_{10})^{-1}$~cm$^{-3}$, an improbably high value. Even when
invoking
turbulent amplification of the magnetic field, or compression by a
radiative shock or by the outflow proposed below, it is difficult to
see how such a large change in RM can be produced along such a small
line of sight. A greatly amplified magnetic field would also cause
depolarization resulting from internal Faraday rotation, yet the level
of fractional polarization in region C is not significantly different
from regions A, B or D. We thus ascribe the differences in RM to
variations in fields and ionized gas in the ambient ISM 
along the line of sight.
Variations (and even reversal
of sign) of the RM have certainly been observed across 
other remnants \cite{dm76}.

\subsubsection{Morphology}
\label{sec_g309_discuss_morph}

The shell and ear components of \snr\ have 
comparable \HI\ absorption spectra and (at least for the north-eastern
ear) similar rotation measures, and are of comparable angular size.
Also, the shell component cannot be seen at the position angles of the
ears.  This leads us to conclude that we are unlikely to be seeing two
structures superimposed, and in further discussion we assume \snr\ to
be a single object.

SNRs with multiple loops and/or significant variations in radius and
brightness are reasonably common -- a perusal of the catalogue of
Whiteoak \& Green \shortcite{wg96}
suggests that $\sim$20 per cent of SNRs have some
sort of multi-ringed structure, corresponding to $\sim45$ Galactic
SNRs of the 215 currently identified \cite{gre96b}. Such structures are
usually interpreted in terms of expansion into multiple cavities in the
ISM \cite{bs86,plr87,mck+89b,dggw94}.  It is possible that
\snr\ represents such a system: the shell component alone is quite
undistorted and would represent the original SNR, while the ears would
trace an ellipsoidal cavity into which the blast wave has then
expanded. However the centres of the shell and ear components coincide
to within 30~arcsec. For two interlocking cavities, we estimate the
probability of such close alignment of their centres to the
line of sight to be $3\times10^{-3}$. Thus even
if all multi-ringed SNRs are due to multiple cavities, it is
unlikely that even one such remnant would show as good an alignment as
seen for \snr. We therefore consider it unlikely that the SNR's
appearance can be explained by simply invoking the inhomogeneity 
of the ISM.

The striking symmetry of the two ears in terms of brightness, shape and
opposed positions around the shell suggest that the mechanism
responsible for them has a characteristic axis. For example, an SNR
evolving in a strong ambient magnetic field will become considerably
elongated along the field direction \cite{ir91,rt95}, perhaps
resembling the ears of \snr. However such a model cannot explain the
round and symmetric shell component, and in any case requires an
ambient field 1000 times stronger than encountered in the ambient ISM.
Another possibility is that the axis is defined by the progenitor
star:  Blondin, Lundqvist \& Chevalier \shortcite{blc96} show that
expansion of a SNR into a progenitor wind of axisymmetric density
distribution causes the shell to develop opposed protuberances,
producing a morphology quite like that of \snr. However as we have
discussed in Paper~I\nocite{gmg98}, this shaping occurs when the
remnant is very young, and it is not clear how the SNR might `remember'
the effects of its progenitor wind as it expands to large sizes.

The alternatives we have just discussed cannot be ruled out, but we argue
that the appearance of \snr\ can be best explained if the SNR contains
a central source which produces collimated outflows or jets in two
opposed directions.  The shell component represents the original,
undistorted SNR. The outflows collide with the expanding remnant, their
pressure distorting and brightening the two opposite sides and thus
producing the ears. \snr\ is then similar to SNR~G039.7--02.0 (W~50),
in which the centrally located X-ray binary SS~433 generates opposed
jets which distort the surrounding shell \cite{eb87,ms96}.  \snr\ is
much younger than W~50 (age $50-100 \times 10^3$~yr), and indeed a
rough progression in morphology can be seen from the former to the
latter.  \snr\ has a distinct, almost circular shell with ears
protruding $\sim$ 30 per cent of the radius beyond it; in W~50,
however, the original shell is much fainter and filamentary, with ears
extending up to twice the radius of the shell from the remnant's centre
\cite{eb87}.

Outflows such as those claimed here are generally associated with jets
produced by an accreting binary system \cite{hj88,fbw97}, but can
also be produced by an isolated neutron star
\cite{sl90,hss+95,bel97}.  An obvious candidate for such a source in
SNR~\snr\ is \src\ (Fig~\ref{fig_g309_snr_central}).  No X-ray
counterpart to \src\ is apparent, and examination of the Digitized Sky
Survey shows no optical source within 8 arcsec.  While the source's
radio morphology is suggestive of the episodic ejections seen in the
X-ray binaries GRS~1915+10 \cite{mr94c} and SS~433 \cite{vss+93}, the
axis so defined is not aligned with the axis for outflow implied by the
SNR morphology, nor does the source itself lie on this symmetry axis.
We thus think it most likely that this source is not associated with
the SNR.

We put a 5-$\sigma$ upper limit of 0.4~mJy on the 1.3~GHz flux density
of any other central source.  No X-ray counterpart is apparent either;
while 1WGA~J1346.5--6255 is suggestive, it
is spatially coincident with and is probably associated with one or
more stars in NGC~5281. The uncertainty in position, the low Galactic
latitude and the proximity to NGC~5281 all make optical identification
difficult.  We thus find no observational evidence for the postulated
central source. However the radio upper limit is consistent with a
binary system in a quiescent state (cf.\ GRO~J1655--40; Hjellming
1997\nocite{hje97}) or an isolated pulsar which is beaming away from
us, while the lack of X-rays can be attributed to absorption in the
Galactic Plane, confusion with diffuse soft emission from the SNR, and
the comparatively short exposure time.
X-ray observations at higher energies and with greater sensitivity with
{\em ASCA}\ or {\em AXAF}\ may
be more successful (cf. Gotthelf, Petre
\& Hwang 1997\nocite{gph97}).

The jet feature seen in our radio image aligns with the axis of
symmetry defined by the two ears, and can be traced back faintly to the
centre of the SNR.  Thus it can be argued that this structure (and
its less distinct counterpart in the remnant's south-west) is indeed emission
representing or surrounding a jet,
and that it delineates the proposed outflow. That the brightest
emission from the jet is produced along a segment just within the
original SNR shell may be due to a sudden change in the jet's
environment, may represent episodic ejections of material, or may
demarcate the progress of a reverse shock generated when the outflow
collides with the SNR shell \cite{ms96}.  There is no evidence for an
outflow in X-rays comparable to the X-ray lobes seen in the interior of
W~50 \cite{sgsg80,wwgs83,yka94}.  However, the {\em ROSAT}\ observations
described in Section~\ref{sec_g309_other_xrays} extend only up to 2.4~keV,
so that absorption along the line of sight may prevent detection.  As
for the central source, hard X-ray observations may overcome this problem.

Comparison of our observations with the 843~MHz image of Whiteoak \&
Green \shortcite{wg96} shows no evidence that the ears have a different
spectral index to the rest of the shell.  The lack of sufficient
separation in frequency betweem the 843~MHz and our observations
prevents a sensitive analysis of spatial spectral index variations
(e.g.\ Anderson \& Rudnick 1993\nocite{ar93}; Gaensler et al.
1998b\nocite{gbm+98}). However, we note that
the ears contribute $\sim$50 per cent of the total flux density of
SNR~\snr: thus if we demand the ears to have a significantly flatter
spectral index than the mean for the SNR, the rest of the remnant must
be steeper by the same amount. No such effect is apparent in the data, 
and it is thus unlikely that the brightening observed in the ears is due to
an injection of relativistic particles from the central outflow. 

The distinct break in the north-eastern ear at the point where it is
crossed by the jet suggests that the pressure of the outflow carries
emitting particles away from the point of impact, thus producing the
apparent break in emission.  Faint emission seen along the axis of the
jet, just outside the north-eastern ear, may represent emitting
electrons which were once part of the shell.

It is interesting to note that the brightest region of the
north-eastern ear is immediately adjacent to the break; a similar
situation exists for the south-western ear.  This may be a result of
additional shocks and turbulence driven into the shocked region of the
SNR by the outflow.  The resulting enhanced particle acceleration in
these regions then produces the bright ears (cf.\ Frail et al.
1997\nocite{fbmo97}).

The column appears to attach to the SNR at the break in the
north-eastern ear, and may join with the jet in the remnant's
interior.  The bends and turns along the column make it difficult to
see how it might be a direct extension of the jet, but we note that
radio emission associated with the possible jet in Vela~X has a
similarly twisted appearance \cite{fbmo97}.  One possibility is that
the column could represent an interaction between the jet and a
distorted old star trail \cite{nl95}. This can explain the bends seen
along the length of the column, and also why no such structure is seen
beyond the opposite ear.

The column broadens at its northern end to
overlap with RCW~80 at the apex of its arc, suggesting that it joins onto
RCW~80.  If the column is unassociated with the SNR, one could interpet
it and RCW~80 as being a single, complex thermal region.  The lack of
optical and infrared emission associated with the column might then be
explained by absorption due to associated molecular material.

On the other hand, if the column is associated with both the SNR and
the \HII\ region, then we have a single system, where the outflow and/or
column connects the SNR with the thermal arc.
Associating \snr\ with RCW~80 puts the SNR at the lower limit of
the distance range inferred from its \HI\ absorption, and implies a
shell radius of $9\pm3$~pc. Assuming expansion into a homogenous medium
of density $n_0$~cm$^{-3}$, we find that the remnant has swept up
between 20 and 170~$M_{\sun}$, and is thus most likely in a transition
between free expansion and the subsequent Sedov-Taylor phase. Assuming
$n_0 = 0.2$ and \mbox{$E_{51} = 1$} as adopted in Paper~I (where
$E_{51}$ is the kinetic energy of the explosion in units of
$10^{51}$~erg), we derive an upper limit on the remnant's age of
4000~yr.

\subsection{Other SNRs with jets}

\mbox{\snr\ bears} a remarkable resemblance to G332.4+00.1 (Kes~32), a roughly
circular shell which is distorted and brightened at one end. A
flat-spectrum collimated structure extends through a break in the shell
in the region of greatest distortion, broadening and kinking before
terminating well outside the remnant in an extended thermal `plume'
\cite{rmk+85,kcr+87}. Higher resolution ATCA observations of
G332.4+00.1 (M.~J.~Kesteven 1997, private communication) show a
collimated feature within the shell near the break.  Thus both
\snr\ and G332.4+00.1 have an undistorted shell component, a distorted
and brightened ear component containing both a jet feature and a break
in the shell, and a winding column which attaches the SNR to a thermal
nebula. The main difference between the two SNRs is that \snr\ appears
to be affected by twin outflows, while G332.4+00.1 seems to involve
only a one-sided jet. The `column' component of \snr\ may have a
steeper spectrum than the equivalent component of G332.4+00.1, but the
uncertainties in the former's spectral index are large because of its
low flux density.

\snr\ also has some resemblance to G320.4--01.2 (MSH~15-5{\em 2}),
X-ray observations of which show a one-sided synchrotron jet emanating
from the central pulsar B1509--58, which then appears to collide with
the H$\alpha$ nebula RCW~89 \cite{tkyb96,bb97,gbm+98}.  As for
\snr\ and G332.4+00.1, the radio morphology of the remnant is
significantly distorted where the outflow intersects the shell.

\begin{table}
\caption{SNRs proposed to contain jet/shell interactions.
`X' and `R'  refer to the existence of X-ray and radio jets
respectively. A reasonable case can be made for the first
four SNRs, while the remaining sources are more speculative.}
\label{tab_g309_outflow}
\begin{tabular}{llcc} 
SNR           & Other name  & Jets? & Reference \\ \hline
G039.7--02.0  & W~50 / SS~433 & XR    & 1, 2 \\
G309.2--00.6  &             & R   &  This paper \\
G320.4--01.2  & MSH~15--5{\em 2} & X & 3, 4, 5 \\
G332.4+00.1   & Kes 32, MSH~16--5{\em 1} & R   &  6, 7 \\ 
\\
0540--693     & Hen~N~158A     & R? & 8 \\
G109.1--01.0  & CTB~109 & X & 9 \\
G290.1--01.8  & MSH 11--6{\em 1}A & X? & This paper \\
G308.8--00.1  &          & -- & 10 \\
G315.9--00.0  &             & R    &  7 \\
G327.6+14.6   & SN~1006     &  --    & 11 \\ \hline
\end{tabular}
\footnotesize
(1)~Seward et al. \shortcite{sgsg80}
(2)~Hjellming \& Johnston \shortcite{hj81a}
(3)~Tamura et al. \shortcite{tkyb96}
(4)~Brazier \& Becker \shortcite{bb97}
(5)~Gaensler et al. \shortcite{gbm+98}
(6)~Roger et al. \shortcite{rmk+85}
(7)~Kesteven et al. \shortcite{kcr+87}
(8)~Manchester et al. \shortcite{msk93}
(9)~Gregory \& Fahlman \shortcite{gf83}
(10)~Kaspi et al. \shortcite{kmj+92}
(11)~Willingale et al. \shortcite{wwps96}
\end{table}

We summarise the SNRs which we have argued are similar to \snr\ in the
first half of Table~\ref{tab_g309_outflow}.
In  the second half of Table~\ref{tab_g309_outflow} we suggest some further
candidates for such interactions. While somewhat speculative, they each
satisfy at least some of the morphological criteria observed in
\snr\ and in the other remnants discussed above. We now briefly discuss
these additional candidates:

\begin{description}

\item {\em 0540--693 (Hen~N~158A)}: This young ($\sim$1000 yr)
composite SNR is in the Large Magellanic Cloud, and is associated with
the Crab-like pulsar PSR~B0540--69.  The shell component of this
remnant is distinctly brightest in the east in  both radio \cite{msk93}
and X-ray \cite{sh94} images, with possible connecting structure to the
central pulsar and its associated synchrotron nebula.  Manchester et
al. \shortcite{msk93} propose that injection of particles from the
pulsar into the shell may cause the observed morphology.

\item {\em G109.1--01.0 (CTB~109)}: This SNR is associated with the
long-period, possibly accreting pulsar 1E~2259+586. Gregory \& Fahlman
\shortcite{gf83} propose that a collimated X-ray feature and the
unusual radio morphology are consistent with precessing jets
originating from the pulsar. More recent observations (e.g.\ Rho
\& Petre 1997\nocite{rp97}) do not
support this interpretation, however.

\item {\em G290.1--01.8 (MSH~11--6{\em 1}A)}: A radio image of this
remnant \cite{wg96} shows it to have two opposed radio
lobes protruding approximately 50 per cent of the SNR radius beyond its
otherwise circular shell, resembling
the `ears' observed in W~50 and claimed here for \snr. We thus propose
this SNR as a possible new example of a shell SNR with jets.  A
distance to G290.1--01.8 of 7~kpc \cite{rho95,ralm96} corresponds to an age
$\sim$10\,000~yr \cite{rho95}, similar to that of \snr. In X-rays, an {\em
ASCA} spectrum of  G290.1--01.8 shows emission lines characteristic
of hot plasma in the remnant's interior \cite{rho95}. However,
there is the suggestion in an earlier {\em Einstein} image of
faint extensions along the axis defined by the lobes \cite{sew90}, and we
speculate that these may be possible analogues of the X-ray jets seen
in W~50.  We note that the inclusion of G290.1--01.8 in this class is
mutually exclusive with a tentative association between it and the
pulsar PSR~J1105--6107, $\sim20'$ from the remnant's centre
\cite{kbm+97}.  Demonstration of either hypothesis would conclusively
rule out the other.

\item {\em G308.8--00.1}: This SNR consists
of two opposed arcs of completely different morphology. The
northern arc has a bright, possibly filled-centre appearance,
which  Kaspi et al. \shortcite{kmj+92}
suggest may be due to an outflow from
the associated pulsar PSR~B1338--62.

\item {\em G315.9--00.0}: This faint shell has a narrow
collimated protrusion extending well beyond its boundary \cite{kcr+87}.
The shell is brightest where the protrusion joins onto it.

\item {\em G327.6+14.6 (SN~1006)}: The remnant of SN~1006 is a classic
barrel SNR. Willingale et al. \shortcite{wwps96} argue that its
symmetric, bilateral appearance  is produced by twin electron beams
originating from an unseen central engine.  We note, however, that many
other explanations have been proposed \cite{kc87,rmk+88,rey96,gae98}.

\end{description}

There thus may be as many as ten SNRs for which
jets from a central source interact with, brighten and/or distort the
surrounding shell. Along with SNRs in which an associated pulsar or
other source is detected directly \cite{kas96,bj97}, there is now a large
and diverse collection of SNRs for which the presence of  a compact
stellar remnant can be {\em inferred}:  there are SNRs affected by jets and
outflows as discussed above, remnants which are distorted and
re-energised by the passage of an associated pulsar through the shell
\cite{sfs89} and SNRs containing traditional pulsar-powered components
(`plerions'; e.g.\ Helfand \& Becker 1987\nocite{hb87}). Evidence that
plerions may also be powered by jets \cite{hss+95,bb97} suggests that the
various interactions between a SNR and a compact source may all be
manifestations of the same phenomenon, the differences depending on the
details of the environment and position of the source within the SNR.

The fraction of SNRs in which a central source is observed or can be
inferred is still well below the $\sim$80\% of supernovae that are of
type~Ib or II \cite{vt91} and thus expected to produce such sources.
This discrepancy has traditionally been explained in terms of a
population of radio pulsars of low luminosity or which are not beaming
towards us \cite{mt77,ts86,man87}. However, we are now aware of a
variety of new complications which prevent such detections:

\begin{enumerate}

\item Observational selection effects may prevent the detection of a
pulsar or its associated nebula. As mentioned above luminosity and
beaming may account for a lack of associations, while emission from a
pulsar nebula may be swamped by emission from the surrounding shell or
from other sources.  An instance in which several of these effects are
operating simultaneously is PSR~B1853+01 in the SNR~G034.7--00.4
(W~44): the pulsar is particularly faint \cite{wcdb88}, and its
surrounding nebula is difficult to distinguish against the rest of the
SNR in radio \cite{fggd96} and in X-rays \cite{hhh96}. 
SN\,1987A may be another object in which emission from the remnant
prevents detection of any central neutron star \cite{mcc93},
while we have argued above that SNR~\snr\ also falls into this category;

\item The environment surrounding a neutron star may prevent
production of a detectable nebula and/or jets. For example,
Bhattacharya \shortcite{bha90} has argued that the lack of a radio
plerion around PSR~B1509--58 can be explained if its progenitor
supernova exploded into a low-density cavity;

\item A compact stellar remnant may be given a high spatial velocity by
its supernova explosion. Once such a source passes well beyond its
SNR's boundaries, it may not be identified as associated with the
remnant, and will probably no longer influence the remnant's morphology
(e.g.\ Gaensler \& Johnston 1995\nocite{gj95a}). A possible example of
such a system is the X-ray binary Circinus X-1, which has been
associated with the nearby SNR~G321.9--00.3 \cite{cpc75,schn93};

\item Some compact remnants may not produce detectable emission and/or
may not interact appreciably with their environment.  Examples include
`injected' neutron stars, born spinning slowly, and `magnetars',
neutron stars born with high magnetic fields which cause them to slow
down at a rapid rate (see Frail 1998\nocite{fra98} for a review and
discussion).  Such sources may not emit radio pulses \cite{cr93a,bh98},
and their nebulae will rapidly fade below detectability \cite{bha90}.
An uncertain fraction of supernovae will produce isolated black holes
\cite{bb94b}, which are also undetectable and which are expected to
show no interaction with their remnant.

\end{enumerate}

\section{Conclusion}
\label{sec_g309_conclusion}

We have presented \HI\ and 1.3-GHz continuum 
observations of SNR~\snr, as well as 
H$\alpha$, {\em ROSAT}\ PSPC and {\em IRAS}\
60~$\mu$m data on the region.
We put a lower limit on linear polarization from the SNR of 1.4 per
cent, a low level which we attribute to beam depolarization. We find a
rotation measure towards most of the SNR of --930~rad~m$^{-2}$, but
a distinctly different RM of \mbox{--570~rad~m$^{-2}$} towards one
component.  This difference is best explained in terms of
ISM differences rather than by conditions
within the SNR itself.  \HI\ absorption puts lower and upper limits on
the SNR's systemic velocity of --50 and +40~\kms\ respectively, putting
it at a distance between 5.4$\pm$1.6 and 14.1$\pm$0.7~kpc and
implying an age in the range $1-20\times10^3$~yr.  The nearby
\HII\ region G309.548--00.737 shows absorption out to the tangent
point, consistent with its recombination line velocity and putting it
at a distance 5.4$\pm$1.6~kpc.

SNR~\snr\ appears to be a typical shell SNR but with two brightened and
distorted `ears' at opposed position angles, which have a similar
spectral index to the rest of the shell.  No emission corresponding to
the remnant is apparent in the infrared or in H$\alpha$, while diffuse
emission can be seen in X-rays. The compact X-ray source
1WGA~J1346.5--6255 within the SNR is probably associated with the
foreground open cluster NGC~5281.

We consider various explanations for the morphology of SNR~\snr, and
argue that the remnant's
appearance is best explained by the presence of
opposed jets from a central source which collide with and distort the
surrounding shell. We propose \snr\ as a possible younger analogue to
the X-ray binary SS~433 and its associated
SNR~W~50.  A faint jet-like structure
oriented along the symmetry axis of \snr\ may correspond to the outflow itself,
while  breaks in the ears along this axis may represent
this outflow travelling beyond the shell. The weak source \src\ in the
SNR's interior is unlikely to be associated with the remnant. We do not
detect any other central source in either X-rays or in radio. 
The former can be attributed to a lack of sensitivity in the observations
and to absorption along the line of sight, while the latter
may indicate a binary system in a quiescent state or a pulsar with
radio beams directed away from us.

To the SNR's north is an unusual column of radio emission, which at one
end may connect with the proposed outflow from the SNR's centre and, at
the other end, with the \HII\ region RCW~80.  Such an association puts
the SNR at a distance 5.4$\pm$1.6~kpc and corresponds to an age of less
than 4000~yr.  The details of the physical process behind such an
interaction are unclear, but we note that a similar combination of
outflow, distortion and termination in a thermal region has been
claimed for both G332.4+00.1 (Kes~32) and G320.4--01.2 (MSH
15--5{\em2}).

Further observations of SNR~\snr\ will be required to determine whether
our interpretation for its appearance is valid. Higher frequency radio
observations can be used to provide higher resolution images of the
`ear' and `jet' regions and the interaction
between them, to better study the polarimetric
properties of the SNR and, together with lower frequency data, to
better constrain any spectral index differences between the different
components of the remnant. If \snr\ is similar to W~50,
X-ray observations of greater sensitivity  and at higher energies
should be able to detect both a central source and evidence for outflow
from it.

Apart from \snr, we find at least eight other SNRs in which the
shell may be affected in some way by jets or outflows
from an associated compact source, and suggest G290.1--01.8
(MSH~11--6{\em 1}A) as a possible further example. While the
characteristic morphology associated with such outflow may become
another means of determining which supernovae have massive star
progenitors, there is good reason to believe that
a significant fraction of SNRs harbour compact remnants
which, for various reasons, we still have not detected.

\section*{Acknowledgments}

We are particularly grateful to Andrew Walker for carrying out the
optical observations and reduction. We also thank Karen Brazier, Jim
Caswell, Mike Kesteven, 
Neil Killeen, Vince McIntyre and Jessica Try for useful
discussions, Simon Johnston for reading the manuscript, Veta Avedisova
for supplying us with information from her catalogue of star formation
regions, and the referee, Tom Landecker, for helpful comments which
improved the paper.  BMG acknowledges the support of an Australian
Postgraduate Award.  The Australia Telescope is funded by the
Commonwealth of Australia for operation as a National Facility managed
by CSIRO. This research has made use of the NASA Astrophysics Data
System, the CDS SIMBAD database, the IPAC HIRES facility and the
HEASARC Online Service, provided by the NASA/Goddard Space Flight
Center.

\bibliographystyle{mn}
\bibliography{modrefs,psrrefs}
 
\label{lastpage}
\end{document}